\documentclass[pdftex,12pt]{article}

\usepackage{amsmath}
\usepackage{amsfonts}
\usepackage{amssymb}
\usepackage{amsthm}
\usepackage{physics}
\usepackage{appendix}
\usepackage{graphicx}
\usepackage{color}
\usepackage[dvipsnames]{xcolor}
\usepackage{cite}

\usepackage{geometry} 
\usepackage{enumitem}
\setenumerate{topsep=3pt,itemsep=2pt}
\setitemize{topsep=3pt,itemsep=2pt}

\usepackage[colorlinks=true,linkcolor=blue,citecolor=magenta,linktocpage=true]{hyperref}

\newcommand{\nextline}{\nonumber \\ &\hspace{1.2em}}
\renewcommand{\:}{\hspace{0.10em}}

\usepackage{bbm}
\newcommand{\one}{\mathbbm{1}}

\newcommand{\pr}[1]{\left(#1\right)}
\newcommand{\pd}{\partial}

\newcommand{\Z}{\mathbb{Z}}
\newcommand{\R}{\mathbb{R}}

\newcommand{\N}{\mathbb{N}}
\newcommand{\DS}{\mathcal{D}}
\newcommand{\HS}{\mathcal{H}}
\newcommand{\LS}{\mathcal{L}}
\newcommand{\MS}{\mathcal{M}}
\newcommand{\VS}{\mathcal{V}}
\newcommand{\ZS}{\mathcal{Z}}

\usepackage{simpler-wick}
\newcommand{\php}{{\phantom{.}}}
\newcommand{\zer}{{\circ}}

\newcommand{\tp}{\tilde{p}}
\newcommand{\tx}{\tilde{x}}
\newcommand{\tVS}{\tilde{\VS}}
\newcommand{\tQ}{\tilde{Q}}
\newcommand{\tq}{\tilde{q}}

\begin{document}

\title{Non-local Field Theory from Matrix Models}

\author{
Andrzej Banburski$^1{}^2$\thanks{t-abanburski@microsoft.com} ,
Jaron Lanier$^1$\thanks{jalani@microsoft.com} ,
Vasudev Shyam$^3$\thanks{vshyam@stanford.edu} , \\
Lee Smolin$^4$\thanks{lsmolin@perimeterinstitute.ca} ,
Yigit Yargic$^1$\thanks{v-yyargic@microsoft.com}
}

\date{
$^1$Microsoft Research, Redmond WA 98052, USA \\[2pt]
$^2$Massachusetts Institute of Technology, Cambridge MA 02139, USA \\[2pt]
$^3$Stanford Institute for Theoretical Physics, Stanford CA 94305, USA \\[2pt]
$^4$Perimeter Institute for Theoretical Physics,
Waterloo ON N2J 2Y5, Canada
\\[3ex] June 27, 2022
}

\maketitle

\begin{abstract}
We show that a class of matrix theories can be understood as an extension of quantum field theory which has non-local interactions. This reformulation is based on the Wigner-Weyl transformation, and the interactions take the form of Moyal $\star$-product on a doubled geometry. We recover local dynamics on the spacetime as a low-energy limit. This framework opens up the possibility for studying novel high-energy phenomena, including the unification of gauge and geometric symmetries in a gauge theory. 
\end{abstract}

\newpage

\tableofcontents


\section{Introduction}

Matrix theories have been motivated through various different approaches in high-energy physics \cite{Banks:1996vh, Ishibashi:1996xs, Taylor:2001vb, Smolin:2000fr, Smolin:2008pk, DiFrancesco:1993cyw, Ambjorn:2008jf, Eichhorn:2013isa, Alexander:2021rch}. Gross-Kitazawa \cite{Gross82} and Kazakov \cite{Kazakov:2000ar} pointed out an interesting relationship between the matrix indices and the momentum-space geometry of a quantum field theory (QFT), in which each matrix entry at index $(j,k)$ is identified as a field degree of freedom at momentum $p=j-k$. The essential property of this representation is that it is preserved under the matrix multiplication, which corresponds to a convolution in $p$. Using this algebraic property, one can express the classical action for a QFT as a pure matrix theory \cite{Yargic:2022ycw}, where the spacetime dependence of fields is replaced by the index dependence of matrix entries.

From the perspective of understanding the matrix dynamics more generally, however, the prescription $p = j-k$ is incomplete. The number of independent variables inside a generic $N \times N$ matrix scales by $N^2$, whereas the number of distinct index differences $j-k$ scales by $N$. Therefore, the spacetime geometry that emerges from $p = j-k$ is only a square-root-sized quotient of an extended background geometry that is intrinsic to large matrices and their algebra. Two distinct matrix entries at $(j,k) \neq (j',k')$ with $j-k = j'-k'$ should correspond in the QFT picture to two degrees of freedom that live on the same spacetime point, but at different points in an auxiliary background structure. Consequently, the geometry of large-$N$ matrix theories is a doubling of the geometry of local QFTs.

In this paper, we explore a geometric understanding of infinite-sized matrix theories. We focus on the case of continuous matrix indices, which corresponds to replacing discrete momenta by continuous ones. To be precise, the `matrices' here are viewed as linear operators on $L^2(\R^d)$. With a Wigner-Weyl transformation \cite{Wigner32, Weyl27}, a matrix $A$ is turned into a field $A(x,\tx)$ with two independent position arguments. The Fourier conjugate of these position arguments are two momentum arguments, $p$ and $\tp$. The matrix multiplication is equivalent in this formalism to the Moyal $\star$-product \cite{Groenewold46, Moyal49} between such fields. The Moyal product is non-local in the arguments $x$ and $\tx$, but conserves $p$ and $\tp$ independently. Thus, from an action principle that consists of variable matrices, we derive an extension to QFT in which the background spacetime is doubled and the interactions are non-local.

There are parallels between our approach to matrix models in this paper and several other lines of research, including deformation quantization \cite{Kontsevich:1997vb}, non-commutative field theory \cite{Doplicher:1994tu, Douglas:2001ba}, double field theory \cite{Hull:2009mi, Aldazabal:2013sca}, and the BFSS \cite{Banks:1996vh} and IKKT models \cite{Ishibashi:1996xs} motivated in string theory and M-theory. Our presentation is self-contained, but we do not intend any disregard to these similar approaches and the vast literature of matrix theories which we could not review here.

Our geometric approach provides a detailed picture for the dominance of planar ribbon diagrams in the perturbative expansion of large-$N$ matrix models \cite{HOOFT1974461, PhysRevLett.48.1063, Marino, Anninos:2020ccj}. As an alternative to the ribbon diagrams carrying two matrix indices, we propose an equivalent expansion with single-line diagrams that carry two conserved momenta. In comparison to a local field theory, the Feynman rules for vertices are changed by a Moyal-type factor that captures the non-local and non-commutative nature of matrix interactions. We show that the different topological embeddings of ribbon diagrams is correctly accounted for by the vertex factors of these single-line diagrams.

A key feature of the models that we discuss here is the presence of a special set of matrices $Q_\mu$ in the action, which commute with each other and provide a background structure. Unlike the variable matrices $A$, which carry the degrees of freedom and are part of the path measure, the matrices $Q_\mu$ are fixed. The interaction between $Q_\mu$ and $A$ releases the first momentum argument $p_\mu$ of each degree of freedom $A(p,\tp)$, therefore giving a dynamics to the variables in the argument $x$. Consequently, $x$ is interpreted as the emerging spacetime coordinate, whereas $\tx$ is interpreted as an independent, secondary coordinate which only contributes to the interactions.

The Wigner-Weyl transformation from a matrix $A$ to a field $A(x,\tx)$ contains a constant parameter $\kappa$ with the mass dimension $-2$. In the limit $\kappa \rightarrow 0$, the matrix interactions become local in $x$ and ultra-local in $\tx$. Then, we recover a local QFT action as the $\kappa \rightarrow 0$ limit of a matrix theory. This is indeed the main message of this paper: \emph{The matrix theories at finite $\kappa$ represent a uniquely defined, non-local, high-energy modification to the interactions in a quantum field theory.}

The paper is organized as follows: We start Section \ref{sec:Framework} by presenting our use of the Wigner-Weyl formalism to reformulate matrix theories as non-local field theories, and we demonstrate this approach with an interacting scalar model. This model is contrasted with the local scalar field theory on the one side, and with a conventional view of the large-$N$ model on the other side. In Section \ref{sec:Gauge}, we discuss the implications of non-Abelian gauge symmetry in this framework. We focus on an extension of Yang-Mills theory, and discover a unification between the gauge and geometric structures of this model. In Section \ref{sec:Geometry}, we discuss the emergence of geometry from the matrix algebra, and the notion of spacetime as a Lagrangian projection. We conclude in Section \ref{sec:Conclusion} by remarking on future directions for this approach to matrix models. We set $\hbar = c = 1$ throughout the paper, and restore them occasionally to highlight the physical units.

\section{Matrix framework}
\label{sec:Framework}

Our goal is to create a geometric picture to understand matrix actions of the form $S = \Tr(f(A))$, where $A$ is an infinite-sized matrix without any spacetime dependence. In the discrete case, a matrix entry can be reparametrized as $A_{j,k} = A_{p,\tx}$ with discrete momentum $p = j-k$ and another discrete parameter $\tx = j+k$. We focus on the case of continuous matrix indices, so that the parameters $p,\tx$ and each of their Fourier conjugates become continuous as well. Our approach is based on the Wigner-Weyl formalism as we discuss in the following.

\subsection{Wigner-Weyl transformation}
\label{sec:Ess}

We consider matrices $A = (A_{q_1,q_2})$ which take continuous row and column indices. Such `matrices' are understood mathematically as linear operators on the Hilbert space $\HS = L^2(\R)$. With a complete and orthonormal basis $\{\ket{q} \in \HS : q \in \R\}$ of $\HS$, we can access the individual matrix entries by $A_{q_1,q_2} \equiv \bra{q_1} A \ket{q_2}$.

In the following, we will reinterpret the matrix entries as the degrees of freedom of a field which lives on a doubled geometry. This geometry is parametrized by the coordinates $(x,\tx)$, or by the conjugate momenta $(p,\tp)$. The matrix indices $q$ will correspond to $q \sim \tx \pm p$ in the geometric picture.

Given any matrix $A$ represented on $\HS$, we turn this matrix into a field $A(x,\tx)$ with the Wigner-Weyl transformation
\begin{align}
\label{Ess0}
    A(x,\tx) = \int \frac{\dd p}{2\pi\kappa} \, e^{ipx} \bra{\tfrac{1}{\kappa} \: \tx + \tfrac{1}{2} \: p} A \ket{\tfrac{1}{\kappa} \: \tx - \tfrac{1}{2} \: p}
    \;.
\end{align}
Intuitively, this transformation reparametrizes the matrix entries by a $45^\circ$-rotation of the row-column grid, followed by a Fourier transformation along one of the axes. We introduced here the constant $\kappa$ which has the mass dimension $-2$.

It will be more convenient to work with these fields $A(p,\tp)$ in the Fourier space\footnote{We denote any function and its Fourier transform with the same symbol, and make the distinction clear from the context.}. The object $A(p,\tp)$ is called the \emph{kernel} \cite{Neumann1931} of the matrix $A$, and it is given by
\begin{align}
\label{Ess2}
    A(p,\tp) = \int \dd q \; e^{-i\kappa q \tp} \bra{q + \tfrac{1}{2} \: p} A \ket{q - \tfrac{1}{2} \: p}
    \;.
\end{align}
This one-to-one map from a matrix $A$ to its kernel $A(p,\tp)$ has the inverse \cite{Weyl27}
\begin{align}
\label{Ess3}
    \bra{q_1} A \ket{q_2} = \kappa \int \frac{\dd\tp}{2\pi} \, e^{i\kappa \: \tp \: (q_1 + q_2)/2} \: A(q_1-q_2, \tp)
    \;.
\end{align}
Given two matrices $A$ and $B$, the kernel of their product $AB$ is equal to the Moyal $\star$-convolution \cite{Neumann1931, Zachos2017} of their kernels,
\begin{align}
\label{Ess4}
    (A B)(p,\tp) = \int \frac{\dd p'}{2\pi} \: \frac{\dd\tp'}{2\pi} \, e^{i \kappa (\tp'p - p'\tp) / 2} \: A(p', \tp') \, B(p-p', \tp-\tp')
    \;.
\end{align}
The factor $e^{i \kappa (\tp'p - p'\tp)/2}$ in this expression \eqref{Ess4} is of central importance to the rest of the paper. It describes the non-local and non-commutative nature of matrix-type interactions at the energy scale $\kappa^{-1/2}$. We also remark on the conservation of both momentum arguments $p,\tp$ on the two sides of \eqref{Ess4}.

To give dynamics to a matrix theory in the parameter $x$, we introduce the fixed matrix $Q$ as
\begin{align}
    Q = \int_{-\infty}^{\infty} \frac{\dd q}{2\pi\kappa} \; q \ketbra{q}{q}
    \;,
\end{align}
which is a continuous version of the diagonal matrix $Q \sim \operatorname{diag}(\ldots, -2, -1, 0, 1, 2, \ldots)$ \cite{Gross82, Kazakov:2000ar}. Its linear spectrum $q \in \R$ gives a uniform density to the emerging geometry. The kernel of $Q$ is the distribution
\begin{align}
\label{Ess5b}
    Q(p,\tp) = (2\pi)^2 \: \frac{i}{\kappa} \, \delta(p) \, \delta'(\tp)
    \;.
\end{align}
Using \eqref{Ess4} and \eqref{Ess5b} with a test matrix $A$, we get
\begin{align}\label{Ess6}\begin{split}
    (QA)(p,\tp) &= \frac{i}{\kappa} \: \frac{\pd}{\pd\tp} \: A(p,\tp) + \frac{1}{2} \: p \: A(p,\tp)
    \;, \\
    (AQ)(p,\tp) &= \frac{i}{\kappa} \: \frac{\pd}{\pd\tp} \: A(p,\tp) - \frac{1}{2} \: p \: A(p,\tp)
    \;.
\end{split}\end{align}
In particular,
\begin{align}
\label{Ess7}
    [Q,A](p,\tp) = p \: A(p,\tp)
    \;.
\end{align}
We observe in \eqref{Ess6} that the presence of the fixed matrix $Q$ in an action would break the translation invariance in $\tx$ unless it only appears inside commutators as in \eqref{Ess7}. When $Q$ appears in commutators, it plays the role of a derivative in the $x$-coordinate.

Finally, the trace of any matrix $L$ is given by the value of its kernel at the origin,
\begin{align}
    \operatorname{Tr}(L)
    = L(p,\tp) \vert_{p = \tp = 0}
    \;,
\end{align}
which is to be viewed as the integral of a Lagrangian density over $x$ and $\tx$.

\subsection{Scalar model}

We demonstrate the framework in Section \ref{sec:Ess} with a simple example. We consider a variable matrix $\varphi$ together with the action
\begin{subequations}\label{KG1}
\begin{align}\label{KG1a}
    S[\varphi] = \operatorname{Tr}\!\bigg(
    \frac{1}{2} \, [iQ, \varphi]^2 - \frac{1}{2} \, m^2 \: \varphi^2 + \frac{\lambda_3}{3!} \: \varphi^3 + \frac{\lambda_4}{4!} \: \varphi^4
    \bigg)
    \;.
\end{align}
In terms of the kernel of $\varphi$, this action takes the form
\begin{align}\label{KG1b}
    S &= \frac{1}{2} \int \dd p \, \dd \tp \; (p^2 - m^2) \, \varphi(p, \tp) \, \varphi(-p, -\tp)
    \nextline
    + \frac{\lambda_3}{3!} \iint \dd p_1 \dd \tp_1 \dd p_2 \dd \tp_2 \, e^{i\kappa (\tp_1 p_2 - p_1 \tp_2)/2} \, \varphi(p_1, \tp_1) \, \varphi(p_2, \tp_2) \, \varphi(-p_1-p_2, -\tp_1-\tp_2)
    \nextline
    + \frac{\lambda_4}{4!} \iiint \dd p_1 \dd \tp_1 \dd p_2 \dd \tp_2 \dd p_3 \dd \tp_3 \, e^{i\kappa (\tp_1 p_2 - p_1 \tp_2 + \tp_2 p_3 - p_2 \tp_3 + \tp_3 p_1 - p_3 \tp_1)/2}
    \nextline \hspace{4em} \times
    \varphi(p_1, \tp_1) \, \varphi(p_2, \tp_2) \, \varphi(p_3, \tp_3) \, \varphi(-p_1-p_2-p_3, -\tp_1-\tp_2-\tp_3)
    \;.
\end{align}
\end{subequations}
We highlight the exponential factors $e^{i \kappa \pr{\cdots\,\,\!\!}}$ in the interaction terms here, which come from \eqref{Ess4}, and are the origin of the physical phenomena that we will discuss. In the low-energy limit $\kappa \rightarrow 0$, we recover from \eqref{KG1b} the scalar field theory action
\begin{align}
\label{KG3}
    S \underset{\kappa \rightarrow 0}{\approx} \int \dd \tx \int \dd x \, \big(
    \tfrac{1}{2} \: (\pd_x \varphi)^2 - \tfrac{1}{2} \: m^2 \varphi^2 + \tfrac{1}{3!} \: \lambda_3 \varphi^3 + \tfrac{1}{4!} \: \lambda_4 \varphi^4
    \big)(x,\tx)
    \;,
\end{align}
Since this action is ultra-local in $\tx$, the measurements for \eqref{KG1} at an energy scale $E \ll \kappa^{-1/2}$ would be blind to the doubling of the underlying geometry, and only measure an approximately local dynamics in the $x$-space.

For the measurements that are sensitive to finite $\kappa$, the physics of the theory \eqref{KG1} differs from the scalar field theory \eqref{KG3}. In particular, the matrix action \eqref{KG1} predicts that the interactions are non-local in $(x, \tx)$. The classical equations of motion for this theory are obtained by varying the action \eqref{KG1} with respect to the matrix $\varphi$, which gives
\begin{align}
    &\frac{\pd^2}{\pd x^2} \, \varphi(x,\tp) = -m^2 \: \varphi(x,\tp) + \frac{\lambda_3}{2} \int \dd\tp' \, \varphi(x-\kappa(\tp-\tp'), \tp') \, \varphi(x+\kappa \tp', \tp-\tp')
    \\ \nonumber &
    +\frac{\lambda_4}{6} \int \dd\tp' \: \dd \tp'' \, \varphi(x-\kappa(\tp-\tp'), \tp') \, \varphi(x-\kappa(\tp-\tp'-\tp''), \tp''-\tp') \, \varphi(x+\kappa \tp'', \tp-\tp'')
    \;.
\end{align}
We wrote this expression here with a partial Fourier transformation from $p$ to $x$ to show the $\tp$-dependent non-locality of the interactions at finite $\kappa$.

\subsection{Feynman rules}
\label{sec:Feynman}

We define these matrix models at the quantum level through perturbation theory. The partition function has a path integral representation
\begin{align}
\label{KG4}
    \ZS = \int \DS\varphi \, e^{iS}
    \;.
\end{align}
The perturbative expansion for the action \eqref{KG1} can be written with line diagrams that carry two conserved charges: the momentum $p$ and a secondary momentum $\tp$. The corresponding Feynman rules are given as follows:
\begin{itemize}
    \item[-] Propagator :
    \hspace{3.3em}
    \includegraphics[scale=0.5]{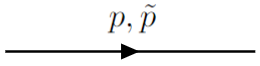}
    \hspace{1.85em}
    $=$ \hspace{0.5em}
    $\dfrac{i}{p^2-m^2}$\;.
    
    \item[-] Cubic vertex :
    \hspace{2em}
    \begin{tabular}{c}
        \includegraphics[scale=0.5]{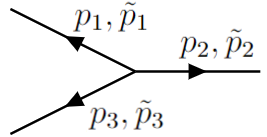}
    \end{tabular}
    \hspace{1.25em}
    $=$ \hspace{0.5em}
    $-i \lambda_3 v_3$\;,
    \begin{align}\label{FeyV3}
        v_3 = \cos(\frac{\kappa}{3} \pr{\tp_1 (p_2 - p_3) + \tp_2 (p_3 - p_1) + \tp_3 (p_1 - p_2)})
        \;.
    \end{align}
    
    \item[-] Quartic vertex :
    \hspace{1em}
    \begin{tabular}{c}
        \includegraphics[scale=0.5]{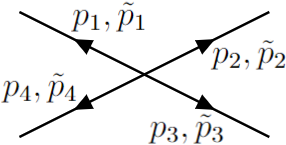}
    \end{tabular}
    \hspace{1em}
    $=$ \hspace{0.5em}
    $-i \lambda_4 v_4$\;,
    \begin{align}\label{FeyV4}
        v_4 = \frac{1}{3} \cos(\frac{\kappa}{2} \pr{(\tp_1 - \tp_3) (p_2 - p_4) - (p_1 - p_3) (\tp_2 - \tp_4)}) + (2 \leftrightarrow 3) + (3 \leftrightarrow 4)
        \;.
    \end{align}
\end{itemize}
A curious property of this model is that the propagator is independent of the secondary momentum $\tp$. We postpone this discussion until Section \ref{sec:Lagr}. The non-trivial effect of $\tp$ appears in the vertices. We note that $v_3,v_4 \rightarrow 1$ in the low-energy limit $\kappa \rightarrow 0$, and we recover the original Feynman rules of scalar field theory. At finite $\kappa$, the coefficients $v_3$ and $v_4$ serve as an ultraviolet correction to the field interactions.

\subsection{\texorpdfstring{Comparison to large-$N$ matrix integral}{Comparison to large-N matrix integral}}

There is an interesting relationship between the matrix model in \eqref{KG1} extending scalar field theory, and the conventional view of the large-$N$ matrix model
\begin{align}
\label{LNMM1}
    \ZS = \int \DS M \, e^{-NS}
    \;, \qquad
    S = \Tr(\tfrac{1}{2} M^2 + \tfrac{\lambda_3}{3!} M^3 + \tfrac{\lambda_4}{4!} M^4)
    \;,
\end{align}
where $M$ is an $N \times N$ matrix. It was shown in \cite{HOOFT1974461} that the perturbative expansion of \eqref{LNMM1} can be written with ribbon diagrams, and the planar diagrams dominate in the limit $N \rightarrow \infty$. We will now give an alternative explanation to this result from the perspective of Wigner-Weyl parameters.

For a matrix theory in the $N \rightarrow \infty$ limit, the matrices would be represented on $\ell^2(\Z)$ rather than $L^2(\R)$. Then, the $p$-space and the $\tx$-space are discrete, whereas the $x$-space and the $\tp$-space are compact. We ignore here the matrix $Q$ in \eqref{KG1a}, and focus solely on the consequences of an infinite volume for the parameter spaces of $p$ and $\tx$.

There are three sources of factors $N$ in the diagrams of the matrix theory \eqref{LNMM1}, which we will denote by $\beta$, $\VS$, $\tVS$ in order to distinguish them: $\beta$ is the coefficient in the exponent in the partition function, $\ZS = \int e^{-\beta S}$. $\VS$ is the volume of the $p$-space (primary momentum). $\tVS$ is the volume of the $\tx$-space (secondary position). We summarize these coefficients in Table \ref{tab:Origins}.

\begin{table}[t]
\renewcommand{\arraystretch}{1.2}
    \centering
    \begin{tabular}{l||c|c|c||}
        & $\beta$ & $\VS$ & $\tVS$ \\ \hline\hline
        Scalar model \eqref{KG1} & $-i/\hbar$ & $\int \dd p$ & $\int \dd \tx$ \\ \hline\hline
        Large-$N$ model \eqref{LNMM1} & $N$ & $N$ & $N$ \\ \hline\hline
    \end{tabular}
    \caption{We distinguish between three different origins of the factors $N$ in the matrix model diagrams.}
    \label{tab:Origins}
\end{table}

These factors appear in the ribbon diagrams for the free energy of a matrix theory as follows:
\begin{itemize}
    \item Each edge gives a factor $\sim \beta^{-1}$.
    \item Each vertex gives a factor $\sim \beta$.
    \item Each loop gives either a factor $\sim \VS$ or a factor $\sim \tVS$ depending on its origin.
\end{itemize}
For the edges in two kinds of diagrams,
\begin{align*}
    \begin{tabular}{c}
        \includegraphics[scale=0.5]{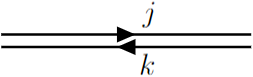}
    \end{tabular}
    \qquad \Leftrightarrow \qquad
    \begin{tabular}{c}
        \includegraphics[scale=0.5]{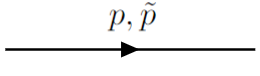}
    \end{tabular}
    \;,
\end{align*}
we identify the parameters with each other based on \eqref{Ess0} as
\begin{align}
    p \sim j-k
    \;, \qquad
    \tx \sim j+k
    \;.
\end{align}
In other words, the primary-momentum $p$ relates to the index difference, and the secondary-momentum $\tp$ relates to the Fourier conjugate of the index sum. We use this identification to determine whether a loop in the ribbon diagrams corresponds to an undetermined $p$ or an undetermined $\tx$.

\subsubsection*{Example 1}

As the first example, we consider the diagrams
\begin{align}
\label{LNMMex1a}
    \begin{tabular}{c}
        \includegraphics[scale=0.4]{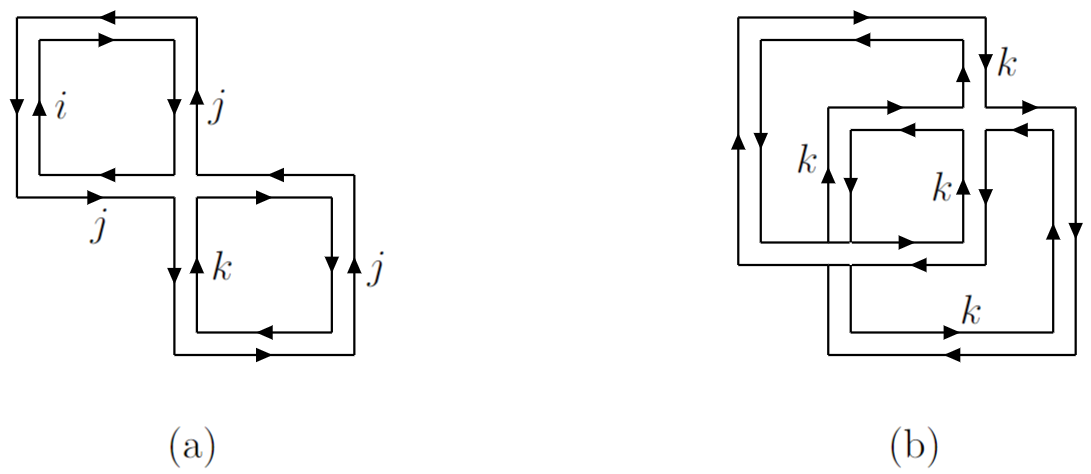}
    \end{tabular}
\end{align}
which come from the contractions $\wick{\c1{M{}^i{}_j} \c1{M{}^j{}_k} \c1{M{}^k{}_l} \c1{M{}^l{}_i}}$ and $\wick{\c1{M{}^i{}_j} \c2{M{}^j{}_k} \c2{M{}^k{}_l} \c1{M{}^l{}_i}}$ for (a), and $\wick{\c1{M{}^i{}_j} \c2{M{}^j{}_k} \c1{M{}^k{}_l} \c2{M{}^l{}_i}}$ for (b). For diagram (a), we have one $\tx$-loop from $(i,j,k) \rightarrow (i+ \Delta \tx, j + \Delta \tx, k + \Delta \tx)$, and two $p$-loops from $i \rightarrow i + \Delta p$ and $k \rightarrow k + \Delta p$. For diagram (b), we have only one $\tx$-loop from $k \rightarrow k + \Delta \tx$. Therefore,
\begin{align}
    \text{(a)} \sim \beta^{-1} \: \VS^2 \: \tVS
    \;, \qquad
    \text{(b)} \sim \beta^{-1} \: \tVS
    \;.
\end{align}
The fact that the planar diagram (a) dominates in the large-$N$ limit is a result of two missing $p$-loops.

We compare this result to the single-line diagram
\begin{align}
\label{LNMMex1c}
    \begin{tabular}{c}
        \includegraphics[scale=0.3]{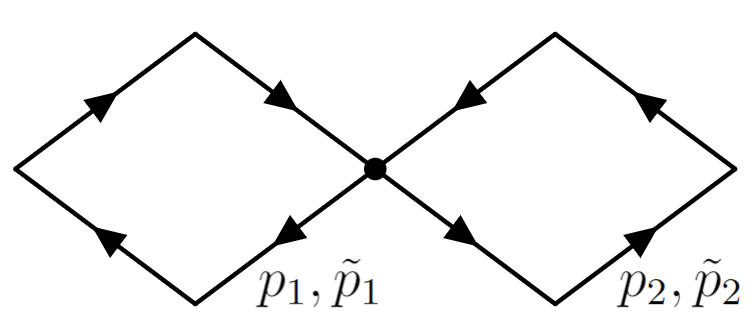}
    \end{tabular}
\end{align}
We claim that the value of this single-line diagram contains the information for both of the ribbon diagrams in \eqref{LNMMex1a} collectively. Recall that we omit $Q$, switch from $L^2(\R)$ to the discrete $\ell^2(\Z) \sim \R^N$, and focus only on the infinite factors $N$ for the sake of this comparison. Based on the Feynman rules from Section \ref{sec:Feynman} and Table \ref{tab:Origins}, the diagram \eqref{LNMMex1c} has a value
\begin{align}
\label{LNMMex1d}
    \sim \tVS \sum_{p_1, p_2} \oint \dd \tp_1 \dd \tp_2 \, \beta^{-1} \beta^{-1} \, \beta \, \tfrac{1}{3} \big(2 + \cos(2\kappa \: (\tp_1 p_2 - p_1 \tp_2))\big)
    \;.
\end{align}
Here, $\tVS$ comes from the fact that this vacuum diagram can take place anywhere in the $\tx$-space\footnote{It can also take place anywhere in the the $x$-space, which has a finite volume in this scenario and thus omitted for the purposes of this discussion.}, and the trigonometric factor $(2 + \cos(\ldots))$ comes from $v_4$ in \eqref{FeyV4} after imposing charge conservation.

The expression \eqref{LNMMex1d} splits into two parts: $\tVS \beta^{-1} \sum_{p_1, p_2} \oint 2 \sim 2 N^2$ corresponds to the planar diagram (a) in \eqref{LNMMex1a}, including the correct symmetry factor of 2 from Wick contractions, whereas $\tVS \beta^{-1} \sum_{p_1, p_2} \oint \cos(\ldots) \sim \tVS \beta^{-1} \delta_{\mathrm{Kron.}}(p_1) \delta_{\mathrm{Kron.}}(p_2) \sim N^0$ corresponds to the non-planar diagram (b) in \eqref{LNMMex1a}. Our claim for consistency between the diagrams \eqref{LNMMex1a} and \eqref{LNMMex1c} is thus confirmed.

\subsubsection*{Example 2}

To give another example, we consider the ribbon diagrams
\begin{align}
\label{LNMMex2a}
    \begin{tabular}{c}
        \includegraphics[scale=0.41]{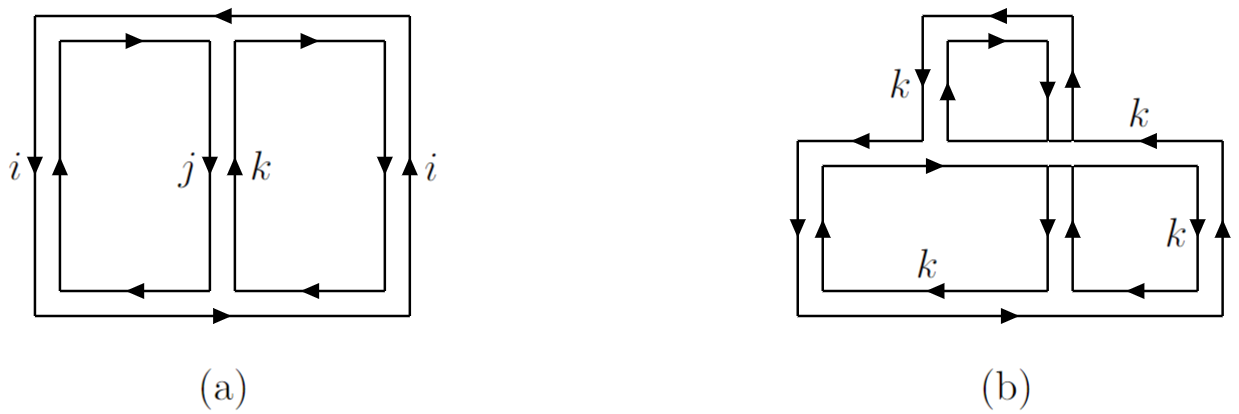}
    \end{tabular}
\end{align}
The diagram (a) has one $\tx$-loop from $(i,j,k) \rightarrow (i+ \Delta \tx, j + \Delta \tx, k + \Delta \tx)$, and two $p$-loops from $j \rightarrow j + \Delta p$ and $k \rightarrow k + \Delta p$. The diagram (b) only has one $\tx$-loop from $k \rightarrow k + \Delta \tx$. Therefore, these diagrams scale with $N$ as
\begin{align}
    \text{(a)} \sim \beta^{-1} \: \VS^2 \: \tVS
    \;, \qquad
    \text{(b)} \sim \beta^{-1} \: \tVS
    \;.
\end{align}
To compare, we look at the single-line diagram
\begin{align}
\label{LNMMex2c}
    \begin{tabular}{c}
        \includegraphics[scale=0.37]{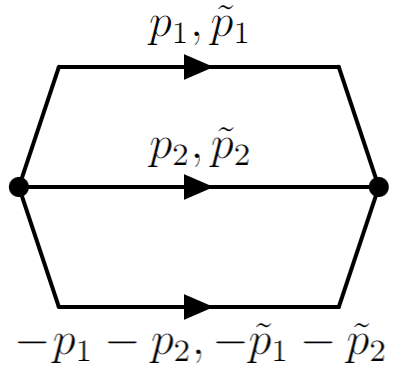}
    \end{tabular}
\end{align}
under the same simplifications as in the previous example. This diagram has a value
\begin{align}
\label{LNMMex2d}
    \sim &\tVS \sum_{p_1, p_2} \oint \dd \tp_1 \dd \tp_2 \, \beta^{-1} \beta^{-1} \beta^{-1} \beta \cos(\kappa \: (\tp_1 p_2 - p_1 \tp_2)) \: \beta \cos(\kappa \: (\tp_1 p_2 - p_1 \tp_2))
    \nonumber \\
    &= \tVS \beta^{-1} \sum_{p_1, p_2} \oint \dd \tp_1 \dd \tp_2 \, \tfrac{1}{2} \big(1+ \cos(2\kappa \: (\tp_1 p_2 - p_1 \tp_2))\big)
    \;.
\end{align}
This expression splits again into two parts: $\tVS \beta^{-1} \sum_{p_1, p_2} \oint 1 \sim N^2$ corresponds to the planar diagram (a), and $\tVS \beta^{-1} \sum_{p_1, p_2} \oint \cos(\ldots) \sim \tVS \beta^{-1} \: \delta_{\mathrm{Kron.}}(p_1) \: \delta_{\mathrm{Kron.}}(p_2) \sim N^0$ corresponds to the non-planar diagram (b) in \eqref{LNMMex2a}. We confirm again that the trigonometric vertex factors from the single-line diagram \eqref{LNMMex2c} are consistent with the dominance of the planar ribbon diagrams at large $N$.

We highlight that these results provide evidence for an equivalence between a topological property of ribbon diagrams and an analytic property of single-line diagrams. While the single-line diagrams in the above examples do not have a non-planar embedding topology as some of the ribbon diagrams they correspond to, their vertex factors \eqref{FeyV3} and \eqref{FeyV4} already contain the relevant information for a sum over ribbon diagrams with different genus.

Finally, we remark on the role of finite $\kappa$ for this large-$N$ phenomenon. If $\kappa = 0$, the cosine terms in \eqref{LNMMex1d} and \eqref{LNMMex2d} would be 1, and the difference in the scaling behavior of planar and non-planar ribbon diagrams would be lost. Hence, this effect of planarity does not carry over to the Feynman diagrams for locally interacting QFTs.

\section{Non-locally interacting gauge theories}
\label{sec:Gauge}

The core premise of analyzing matrix models with the Wigner-Weyl formalism is to deduce a non-local, ultraviolet correction to the locally interacting QFTs, which is uniquely determined by the linear algebra up to a single dimensionful constant $\kappa$. In fact, the gauge symmetries in this context become more restrictive than those in a local QFT, and the theories which remain gauge-invariant become harder to find.

In the following, we first show how the minimal extension of $\mathfrak{su}(n)$ Yang-Mills theory into the matrix framework fails to be gauge-invariant except when $\kappa = 0$. Then we introduce an alternative model, where the gauge symmetry is restored at finite $\kappa$ by including an additional, dynamical component to the gauge boson and replacing the Lie algebra with a finite matrix algebra. This new prototype of gauge symmetries contains a remarkable unification that we explain in Section \ref{sec:gag}.

\subsection{A na\"ive attempt}

To replicate the $\mathfrak{su}(n)$ Yang-Mills theory in $d$ spacetime dimensions, we consider matrices represented on $\HS = L^2(\R^d)$, and we take the fixed matrices
\begin{align}
\label{SYM2}
    Q_\mu = \int_{\R^d} \frac{\dd^dq}{(2\pi\kappa)^d} \; q_\mu \ketbra{q}{q}
    \;, \qquad
    Q_\mu(p,\tp) = (2\pi)^{2d} \, \frac{i}{\kappa} \, \delta(p) \, \frac{\pd}{\pd \tp^\mu} \: \delta(\tp)
    \;.
\end{align}
Note that the subscript of $Q_\mu$ is merely a label for different matrices here, and it does not imply a manifold structure. These matrices commute with each other, $[Q_\mu, Q_\nu] = 0$. For variable matrices $A^a_\mu$, where $a \in \{1,\ldots,n^2-1\}$ and $\mu \in \{1,\ldots,d\}$, we consider the curvature defined in the familiar way as
\begin{align}
\label{YMna2}
    F^a_{\mu\nu} = 2 \: [i Q_{[\mu}^\php, A^a_{\nu]}] + g f^{abc} \: \tfrac{1}{2} \{A^b_{\mu}, A^c_{\nu}\}
    \;,
\end{align}
and the action
\begin{align}
\label{YMna3}
    S = \mathrm{Tr}\big(F^a_{\mu\nu} \: F^{a\mu\nu}\big)
    \;.
\end{align}
Note that \eqref{YMna2} includes a symmetrization between non-commuting matrices which ensures that the curvature is anti-symmetric in the indices $\mu$, $\nu$. However, this action \eqref{YMna3} is not invariant under the gauge transformations
\begin{align}
    \delta A^a_\mu &= \tfrac{1}{g} [iQ_\mu^\php, \alpha^a]
    - f^{abc} \: \tfrac{1}{2} \{\alpha^b, A^c_\mu\}
    \;.
\end{align}
This non-invariance is apparent in the lack of gauge covariance for the curvature \eqref{YMna2}, as we find
\begin{align}
    \delta F^a_{\mu\nu} = - f^{abc} \: \tfrac{1}{2} \{\alpha^b, F^c_{\mu\nu}\} + \tfrac{1}{2} \: g f^{adc} f^{cbe} \: \big[\big[\alpha^b, A^d_{[\mu}\big], A^e_{\nu]}\big]
    \;.
\end{align}
The second term breaks the symmetry. This second term consists of commutators, which implies that it contains factors of $\sin(\kappa (\tp'p-p'\tp))$ when written out using \eqref{Ess4}. Therefore the symmetry would be restored in the local QFT limit $\kappa = 0$. In order to construct a matrix theory with a non-Abelian gauge symmetry at finite $\kappa$, we need to include a final component to the gauge boson.

\subsection{\texorpdfstring{Yang-Mills at finite $\kappa$}{Yang-Mills at finite kappa}}

We denote by $\tau^a$ the traceless, $n \times n$ matrices which generate $\mathfrak{su}(n)$ for some fixed $n \in \N$. These matrices satisfy
\begin{align}
\label{SYM1}
    \tau^a \tau^b = \frac{1}{2} \: h^{abc} \: \tau^c + \frac{1}{2n} \: \delta^{ab} \: \one_n
    \;, \qquad h^{abc} = d^{abc} + i f^{abc}
    \;,
\end{align}
where the structure constants $f^{abc}$ are totally skew-symmetric, and $d^{abc}$ are totally symmetric. We focus on the role of $\tau^a$ as a basis of matrices rather than Lie algebra generators.

Let $A^a_\mu$ and $A^\zer_\mu$ be variable matrices, where $a \in \{1,\ldots,n^2-1\}$, and $\zer$ plays the role of $n^2$-th gauge index. We find a gauge-invariant action for these degrees of freedom that is given by
\begin{subequations}\label{SYM3}\begin{align}
    S &= \Tr(F^a_{\mu\nu} F^{a\mu\nu} + 2n \: F^\zer_{\mu\nu} F^{\zer\mu\nu})
    \;, \\
    F^a_{\mu\nu} &= 2 [i Q_{[\mu}^\php, A^a_{\nu]}] - ig h^{abc} A^b_{[\mu} A^c_{\nu]} - 2ig \: [A^\zer_{[\mu}, A^a_{\nu]}]
    \;, \\
    F^\zer_{\mu\nu} &= 2 [i Q_{[\mu}^\php, A^\zer_{\nu]}] - 2ig A^\zer_{[\mu} A^\zer_{\nu]} - \frac{ig}{n} \: A^a_{[\mu} A^a_{\nu]}
    \;.
\end{align}\end{subequations}
When written in terms of the matrix kernels, this action has the explicit form
\begin{subequations}\label{SYM4}\begin{align}
    S &= \int \dd^dp \, \dd^d\tp \, \big(F^a_{\mu\nu}(p,\tp) \, F^{a\mu\nu}(-p,-\tp) + 2n \, F^\zer_{\mu\nu}(p,\tp) \, F^{\zer\mu\nu}(-p, -\tp)\big)
    \;, \displaybreak[0] \\[0.4em]
    F^a_{\mu\nu}(p,\tp) &= i p_\mu A^a_\nu(p,\tp) - i p_\nu A^a_\mu(p,\tp)
    \nextline
    + g f^{abc} \int \dd^dp' \: \dd^d\tp' \: \cos(\kappa (\tp \: p' - p \: \tp')) \: A^b_\mu(p-p', \tp-\tp') \: A^c_\nu(p',\tp')
    \nextline
    + g \int \dd^dp' \: \dd^d\tp' \: \sin(\kappa (\tp \: p' - p \: \tp')) \, \Big(
    d^{abc} A^b_\mu(p-p', \tp-\tp') \: A^c_\nu(p',\tp')
    \nextline \hspace{13em}
    + 4 A^\zer_{[\mu}(p-p', \tp-\tp') \: A^a_{\nu]}(p',\tp') \Big)
    \;, \displaybreak[0] \\[0.2em]
    F^\zer_{\mu\nu}(p,\tp) &= i p_\mu A^\zer_\nu(p,\tp) - i p_\nu A^\zer_\mu(p,\tp)
    \nextline
    + g \int \dd^dp' \: \dd^d\tp' \: \sin(\kappa (\tp \: p' - p \: \tp')) \, \Big(
    2 A^\zer_\mu(p-p', \tp-\tp') \: A^\zer_\nu(p',\tp')
    \nextline \hspace{13em}
    + \tfrac{1}{n} \: A^a_\mu(p-p', \tp-\tp') \: A^a_\nu(p',\tp') \Big)
    \;.
\end{align}\end{subequations}
Note that the degrees of freedom in $A^a_\mu$ are coupled to those in $A^\zer_\mu$ at finite $\kappa$. In the limit $\kappa \rightarrow 0$, the two sectors are decoupled from each other, and their dynamics becomes ultra-local in the $\tx$-space. A low-energy ($\ll \kappa^{-1/2}$) observer would see $\mathfrak{su}(n)$ Yang-Mills theory plus Maxwell theory propagating in the $x$-space.

This action (\ref{SYM3}, \ref{SYM4}) is invariant under the gauge transformations
\begin{subequations}\label{SYM5}\begin{align}
    \delta A^a_\mu &= \tfrac{1}{g} [i Q_\mu^\php, \alpha^a]
    + \tfrac{i}{2} h^{abc} \: \alpha^b A^c_\mu
    - \tfrac{i}{2} h^{acb} \: A^c_\mu \alpha^b
    + i \: [\alpha^a, A^\zer_\mu]
    + i \: [\alpha^\zer, A^a_\mu]
    \;, \\
    \delta A^\zer_\mu &= \tfrac{1}{g} [i Q_\mu, \alpha^\zer] + i \: [\alpha^\zer, A^\zer_\mu] + \tfrac{i}{2n} \: [\alpha^a, A^a_\mu]
    \;,
\end{align}\end{subequations}
where $\alpha^a$ and $\alpha^\zer$ are arbitrary infinitesimal-valued matrices. Since the kernels of these phases vary in both $x$ and $\tx$, the interpretation of this new type of symmetry \eqref{SYM5} as a `gauge symmetry' is justified. As two parts of the same symmetry group at finite $\kappa$, the gauge components $A^a_\mu$ and $A^\zer_\mu$ have the same coupling constant $g$.

If the symmetry group of the Standard Model is the result of a spontaneous symmetry breaking from an $\mathfrak{su}(n)$ Grand Unified Theory at a higher energy scale, then the matrix model approach further predicts the existence of an accompanying gauge boson $A^\zer_\mu$ which couples to the other gauge bosons at the energy scale $\kappa^{-1/2}$.

We remark that the component $A^\zer_\mu$ in this model parallels the $U(1)$ part of the Yang-Mills matrix model discussed in \cite{Steinacker:2010rh}, and understood as carrying gravitational degrees of freedom. An important difference between the two models is that the matrices $Q_\mu$ here commute with each other. Nonetheless, we speculate that the gravity interpretation might hold for this model as well.

\subsection{Gauge-geometry unification}
\label{sec:gag}

The action \eqref{SYM4} has the gauge symmetry \eqref{SYM5} which is a local symmetry in both $x$ and $\tx$. This symmetry goes beyond the Lie algebra $\mathfrak{su}(n)$, and the theory has the new degrees of freedom $A^\zer_\mu$ compared to its Yang-Mills counterpart. This seemingly complicated structure hides a deeper symmetry principle that unifies its gauge and kinematic aspects.

We write matrices represented on the larger Hilbert space $\HS' = \R^n \otimes L^2(\R^d)$ with a bold font. With
\begin{align}
    \boldsymbol{Q}_\mu = \one_n \otimes Q_\mu
    \qquad \text{and} \qquad
    \boldsymbol{A}_\mu = \one_n \otimes A^\zer_\mu + \tau^a \otimes A^a_\mu
    \;,
\end{align}
the curvature \eqref{SYM3} becomes
\begin{align}
\label{gag2}
    \one_n \otimes F^\zer_{\mu\nu} + \tau^a \otimes F^a_{\mu\nu}
    = \tfrac{i}{g} \: [\boldsymbol{Q}_\mu + g \boldsymbol{A}_\mu, \boldsymbol{Q}_\nu + g \boldsymbol{A}_\nu]
    \;.
\end{align}
In this way, the action \eqref{SYM4} can be written as $S = \Tr([\boldsymbol{Q}_\mu + g \boldsymbol{A}_\mu, \boldsymbol{Q}_\nu + g \boldsymbol{A}_\nu]^2)$, and the gauge transformation \eqref{SYM5} becomes $\delta \boldsymbol{A}_\mu = \frac{i}{g} [\boldsymbol{Q}_\mu + g \boldsymbol{A}_\mu, \boldsymbol{\alpha}]$. We can write a gauge-covariant derivative as $\nabla_\mu \boldsymbol{V} = i [\boldsymbol{Q}_\mu + g \boldsymbol{A}_\mu, \boldsymbol{V}]$. The curvature follows from $[\nabla_\mu, \nabla_\nu] \boldsymbol{V} = ig [\boldsymbol{F}_{\mu\nu}, \boldsymbol{V}]$, and satisfies the Bianchi identity $\nabla_{[\mu} \boldsymbol{F}_{\nu\rho]} = 0$.

To give an intuitive picture, we can think of the Hilbert space $\HS' = \R^n \otimes L^2(\R^d)$ as $\HS' \sim \R^{nN}$, where $N = |\R^d|$ is the cardinality of spacetime. Then, the symmetry algebra of the theory \eqref{SYM4} is simply
\begin{align}
\label{gag3}
    \mathfrak{g} = \mathfrak{su}(nN)
    \;.
\end{align}
What we mean here precisely is that we have a binary operation $\bigstar$ between the $n \times n$-matrix-valued fields on $\R^d \times \R^d$, which combines matrix multiplication with the Moyal $\star$-product. The symmetry algebra \eqref{gag3} is based on the $\bigstar$-commutator
\begin{align}
\label{gag4}
    &(M^j \alpha_j) \bigstar (M^k \beta_k) - (M^k \beta_k) \bigstar (M^j \alpha_j)
    \nonumber \\
    &= \frac{1}{2} \: (M^j M^k - M^k M^j) (\alpha_j \star \beta_k + \beta_k \star \alpha_j) + \frac{1}{2} \: (M^j M^k + M^k M^j) (\alpha_j \star \beta_k - \beta_k \star \alpha_j)
    \,,
\end{align}
where $\{M^j\}$ is a basis of all $n \times n$-matrices. When $\kappa \rightarrow 0$, the $\star$-product becomes commutative. Then, we only observe the first term on the right-hand side of \eqref{gag4}, which turns the symmetry algebra into
\begin{align}
    \mathfrak{g} \,\,\underset{\kappa \rightarrow 0}{\longrightarrow}\,\,\, \mathfrak{u}(1)^{N^2-1} \;\;+\;\; \underbrace{\mathfrak{su}(n) \;\;\times \underbrace{\mathfrak{u}(1)^N}_{\substack{\text{$x$-space} \\ \text{configurations}}}}_{\substack{\text{the standard Yang-Mills} \\ \text{gauge symmetry}}} \times \underbrace{\mathfrak{u}(1)^N}_{\substack{\text{$\tx$-space} \\ \text{configurations}}}
    \;.
\end{align}
At finite $\kappa$, we can write
\begin{align}\begin{split}
    \mathfrak{su}(nN) &\simeq \mathfrak{u}(1) \times \mathfrak{su}(N) + \mathfrak{su}(n) \times \mathfrak{u}(N)
    \;, \\
    \boldsymbol{A}_\mu &= \one_n \otimes A^\zer_\mu + \tau^a \otimes A^a_\mu
    \;,
\end{split}\end{align}
where $\simeq$ is an equivalence of linear spaces. $\mathfrak{su}(N)$ stands for the space of traceless matrices on $L^2(\R^d)$, in the sense $A^\zer_\mu(p,\tp) \vert_{p = \tp = 0} = 0$, because the part of $A^\zer_\mu$ that is proportional to a multiple of identity is non-dynamical as we notice in \eqref{SYM3}.

This picture explains why the symmetry would not close at finite $\kappa$ without the degrees of freedom $A^\zer_\mu$: It is because the space $\mathfrak{su}(n) \times \mathfrak{u}(N)$ is not closed under the $\bigstar$-commutator. Precisely, if we restrict $(M^j \alpha_j)$ and $(M^k \beta_k)$ in \eqref{gag4} such that their basis $\{M^i\} = \{\tau^a\} \cup \{\one_n\}$ only runs over the traceless matrices, the outcome of the $\bigstar$-commutator \eqref{gag4} would not satisfy the same property. Therefore, the $U(1)$-like part $A^\zer_\mu$ of the gauge boson is a necessary component for the gauge symmetry to exist at finite $\kappa$ in this matrix model.

Extending the Yang-Mills gauge symmetry to $\mathfrak{g} = \mathfrak{su}(nN)$ also has another interesting consequence: It puts the gauge $(n)$ and kinematic $(N)$ structures of the theory on an equal footing. As far as the degrees of freedom $\boldsymbol{A}_\mu$ are concerned, the Hilbert space $\HS' \sim \R^{nN}$ does not have any preferred basis, and therefore there is no way to distinguish between the gauge and kinematic parts of $\boldsymbol{A}_\mu$ intrinsically.

The gauge-geometry distinction in this model emerges from the fixed matrix kernel $\boldsymbol{Q}_\mu(p,\tp) = \one_n \frac{i}{\kappa} \: \delta(p) \: \pd_\mu \delta(\tp)$. This object creates a preferred splitting $\R^{nN} \simeq \R^n \times \R^N$, and it makes the degrees of freedom propagate in $\R^N \subset \R^{N^2}$, but not in $\R^{n^2}$.

We conclude this discussion here with an analogy between the proposed gauge-geometry unification at finite $\kappa$ and the space-time unification in special relativity at finite $c$. We presume the constant $\kappa$ might be the Planck scale, $\kappa \approx m_{\text{Planck}}^{-2}$. Similarly to the Newtonian limit $c \rightarrow \infty$ in which the space and the time are seen as distinctly-behaving entities, we propose here that the difference in how we view the gauge and the geometry might be a bias from the low-energy approximation $\kappa \rightarrow 0$ at the measurement scale of the LHC.

\section{Geometry of the kernels}
\label{sec:Geometry}

The basic idea of our approach is to rewrite the matrices in a matrix model as field variables using the Wigner-Weyl transformation, which turns it into a non-local field theory on a doubled geometry. The interaction between any two fields $A(x,\tx)$ and $B(x,\tx)$ is dictated by the matrix multiplication, and it is given by the Moyal $\star$-product \cite{Hillery84}
\begin{align}
\label{geo7}
    (AB)(x,\tx) = \int \dd^d x_1 \: \dd^d\tx_1 \: \dd^d x_2 \: \dd^d\tx_2 \, e^{\frac{i}{\kappa} (x_1 \tx_2 - \tx_1 x_2)} A(x+x_1, \tx+\tx_1) \: B(x+x_2, \tx+\tx_2)
    \,.
\end{align}
In this section, we discuss some of the geometric aspects of these models.

\subsection{Probing the geometry}

The matrices $Q_\mu$ from \eqref{SYM2} have the eigenvalues $q_\mu \sim \tx_\mu \pm p_\mu$, and they can be used in a commutator or in an anti-commutator with a test matrix $A$ as in \eqref{Ess6} to probe the parameters $p$ or $\tx$ for that matrix, respectively.

We consider a complementary set of matrices $\tQ^\mu$ defined with the kernel
\begin{align}
    \tQ^\mu(p,\tp) = - \: (2\pi)^{2d} \: \frac{i}{\kappa} \, \delta(\tp) \, \frac{\pd}{\pd p_\mu} \: \delta(p)
    \;.
\end{align}
The eigenvalues of $\tQ^\mu$ correspond to $\tq^\mu \sim x^\mu \pm \tp^\mu$, therefore it probes the parameters $x$ or $\tp$ for a test matrix $A$ such that
\begin{align}\begin{split}
\label{prob2}
    (\tilde{Q}^\mu A)(p,\tp) &= - \frac{i}{\kappa} \: \frac{\pd}{\pd p_\mu} \: A(p,\tp) + \frac{1}{2} \: \tp^\mu A(p,\tp)
    \;, \\
    (A \tilde{Q}^\mu)(p,\tp) &= - \frac{i}{\kappa} \: \frac{\pd}{\pd p_\mu} \: A(p,\tp) - \frac{1}{2} \: \tp^\mu A(p,\tp)
    \;.
\end{split}\end{align}
The commutator between the matrices $Q_\mu$ and $\tQ^\mu$ is given by
\begin{align}
    [Q_\mu, \tilde{Q}^\nu](p,\tp) = \frac{i}{\kappa^2} \, \sin(\kappa \tp p) \, \pd^\nu \delta(p) \, \pd_\mu \delta(\tp)
    \;.
\end{align}
This distribution can be further simplified by evaluating it on a test matrix $A$ using the Moyal $\star$-convolution \eqref{Ess4}, and we get the canonical commutation relation
\begin{align}
    [Q_\mu, \tilde{Q}^\nu](p, \tp) = \frac{i}{\kappa} \, \delta_\mu^\nu \, \delta(p) \, \delta(\tp)
    \;.
\end{align}
In one dimension, the exponentiated matrices
\begin{align}\begin{split}
    (e^{i\alpha Q})(p,\tp) &= (2\pi)^2 \: \delta(p) \: \delta(\tp - \alpha/\kappa)
    \;, \\
    (e^{i\beta \tQ})(p,\tp) &= (2\pi)^2 \: \delta(p + \beta / \kappa) \: \delta(\tp)
\end{split}\end{align}
are infinite-sized clock and shift matrices \cite{Radicevic:2021ykf}, which form a generalized Clifford algebra by the relation $e^{i\alpha Q} e^{i\beta \tQ} = e^{-i\alpha\beta/\kappa} e^{i\beta \tQ} e^{i\alpha Q}$. They act as translation operators on the $x$-space and $\tx$-space, respectively,
\begin{align}\begin{split}
    (e^{i\alpha Q} A \: e^{-i\alpha Q})(x,\tx) = A(x+\alpha, \tx)
    \;, \\
    (e^{i\beta \tQ} A \: e^{-i\beta \tQ})(x,\tx) = A(x, \tx + \beta)
    \;.
\end{split}\end{align}
Going back to \eqref{Ess6} and \eqref{prob2}, we have four different ways of probing the Wigner-Weyl parameters of a test matrix $A$,
\begin{align}
\begin{array}{lll}
    \displaystyle
    x^\mu A = - \kappa \: \{\tilde{Q}^\mu, A\}
    \;, & \qquad &
    \displaystyle
    p_\mu A = [Q_\mu, A]
    \;,
    \\[0.8em]
    \displaystyle
    \tx_\mu A = \kappa \: \{Q_\mu, A\}
    \;, & \qquad &
    \displaystyle
    \tp^\mu A = [\tilde{Q}^\mu, A]
    \;.
\end{array}
\end{align}
From these relations, we can derive the operational commutator for probing these different parameters in different orders. Firstly, we find
\begin{subequations}\label{geo6}\begin{align}
    [x^\mu, p_\nu] \: A &\equiv
    \{-\kappa \tQ^\mu, [Q_\nu, A]\} - [Q_\nu, \{-\kappa \tQ^\mu, A\}]
    \nonumber \\
    &= - \kappa \: \{[\tilde{Q}^\mu, Q_\nu], A\}
    \nonumber \\
    &= i \delta^\mu_\nu \: A
    \;,
\intertext{and}
    [\tx_\mu, \tp^\nu] \: A &\equiv
    \{ \kappa Q_\mu, [\tQ^\nu, A]\} - [\tQ^\nu, \{\kappa Q_\mu, A\}]
    \nonumber \\
    &= \kappa \: \{[Q_\mu, \tilde{Q}^\nu], A\}
    \nonumber \\
    &= i \delta_\mu^\nu \: A
    \;.
\end{align}\end{subequations}
These equations agree with the quantum mechanical non-commutativity relations $[x,p] = i\hbar \: \one$ and $[\tx,\tp] = i\hbar \: \one$, where $\hbar$ is restored by dimensional analysis. Furthermore, we find
\begin{subequations}\label{geo8}\begin{align}
    [x^\mu, \tx_\nu] \: A &\equiv
    \{ -\kappa \tQ^\mu, \{ \kappa Q_\nu, A\}\} - \{ \kappa Q_\nu, \{ -\kappa \tQ^\mu, A\}\}
    \nonumber \\
    &= - \kappa^2 \: [[\tilde{Q}^\mu, Q_\nu], A]
    \nonumber \\
    &= 0
    \;,
\intertext{and}
    [p_\mu, \tp^\nu] \: A &\equiv
    [Q_\mu, [\tQ^\nu, A]] - [\tQ^\nu, [Q_\mu, A]]
    \nonumber \\
    &= [[Q_\mu, \tilde{Q}^\nu], A]
    \nonumber \\
    &= 0
    \;.
\end{align}\end{subequations}
These results indicate there is \emph{no} uncertainty relation between the coordinates $(x,\tx)$ of the doubled geometry $\MS = \R^{2d}$ that we obtain from the Wigner-Weyl formalism. The non-commutativity in our approach lies in the interaction terms, but not in the background geometry itself. This is because our approach is not set up with fields that take non-commuting arguments, but the coordinate arguments are obtained from a Wigner-Weyl transformation of pure matrices.

\subsection{Lagrangian subspace}
\label{sec:Lagr}

The background $\MS = \R^{2d}$ of the matrix kernels is a commutative symplectic space, where the symplectic structure $\omega$ can be read from the Moyal product \eqref{geo7} as
\begin{align}
    \omega\big((x, \tx), (x', \tx')\big) = x^\mu \: \tx'_\mu - \tx_\mu \: x'^\mu
    \;.
\end{align}
This symplectic structure is an intrinsic property of the linear algebra between infinite-sized matrices. In particular, any symplectic transformation of the dual space $\MS^*$ is manifested uniformly as a general linear transformation on all matrices. For example,
\begin{align}
    \big(e^{i \epsilon \kappa \tQ^2} A \: e^{-i \epsilon \kappa \tQ^2}\big)(p,\tp) = A(p,\tp) + \epsilon \tp \, \frac{\pd}{\pd p} \: A(p,\tp) = A(p + \epsilon \tp, \tp)
    \;.
\end{align}
We observe that the general linear transformations generated by $\tQ^2$ correspond to the symplectic transformations $(p',\tp') = (p + \epsilon \tp, \tp)$ on $\MS^*$. Similarly, the symplectic transformation $(p',\tp') = (p, \tp + \epsilon p)$ is generated by $Q^2$, and $(p',\tp') = ((1+\epsilon) \: p, (1- \epsilon) \: \tp)$ is generated by $Q\tQ$ at the level of matrices. Other matrix similarity transformations generally have a non-linear effect on the coordinates in $\MS^*$.

On any Lagrangian subspace $\LS^* \subset \MS^*$ where the symplectic form $\omega((p,\tp), (p',\tp')) \vert_{\LS^*}$ vanishes, the Moyal $\star$-convolution \eqref{Ess4} becomes an ordinary convolution, and we recover local dynamics on $\LS$. For example, the projection onto the subspace $\tp = 0$ corresponds to $\pd A(x,\tx) / \pd \tx = 0$ for all field variables, or equivalently to constraining the variable matrices to be Toeplitz-type \cite{Yargic:2022ycw}, and we would obtain a local field theory in the spacetime parameter $x$.

A matrix model which only contains generic, variable matrices in its action does not have a preferred Lagrangian subspace. In the two models \eqref{KG1a} and \eqref{SYM3} that we discussed in this paper, the fixed matrix $Q$ is responsible for selecting out a preferred Lagrangian subspace, or equivalently for selecting out a preferred basis of $\HS$ for the Wigner-Weyl transformation \eqref{Ess2}. We have $[Q,A](p,\tp) = p A(p,\tp) \propto \frac{\pd}{\pd x} A$, and the matrix $Q$ introduces a notion of propagation on the $x$-space for these models, thereby distinguishing $x$ from $\tx$.

In particular, we made a deliberate modelling decision in the actions \eqref{KG1} and \eqref{SYM3} by using $Q$ but not also $\tQ$ to create dynamics. The duality between $x$ and $\tx$ is broken by $Q$ in these models. The propagator \includegraphics[scale=0.4]{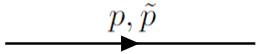} $= \frac{i}{p^2-m^2}$ depends on $p$ but not on $\tp$. Consequently, the parameter space of $\tp$ contributes only to dressing the interactions, as in the vertex Feynman rules \eqref{FeyV3} and \eqref{FeyV4}. We preserve a world-line picture for scattering processes based on the Lagrangian subspace of the parameter $x$, so that these models are better comparable to a conventional QFT. These features can be considered as optional for model-building purposes in the matrix framework more generally.

\section{Outlook}
\label{sec:Conclusion}

We presented a geometric reformulation of infinite-sized matrix theories as non-local field theories. This approach has a predictive potential for the ultraviolet structure of interacting QFTs, but using this potential has some open challenges.

We did not discuss fermions in this paper. While building a free Dirac action for Grassmann-valued matrices would be straightforward, their non-local interaction with the gauge bosons and the possible anomalies of such interactions require further analysis.

The dimensionful constant $\kappa$ determines universally the scale at which all interactions in a matrix theory are non-local. We presume that this constant corresponds to the Planck scale, $\kappa \approx m_{\mathrm{Planck}}^{-2} = G_{\mathrm{Newton}} / \hbar$. Our matrix models recover a locally interacting QFT on a flat spacetime precisely in the limit $\kappa \rightarrow 0$.

An open question is then how one might be able to recover gravity in this approach. Since the fixed matrices $Q_\mu$ in our models determine the Lagrangian subspace that gives rise to the spacetime, a curvature on that subspace would be associated with the degrees of freedom from a variation of $Q_\mu$. In fact, the role of varying $Q_\mu$ is played by the variable component $A^\zer_\mu$ in the finite-$\kappa$ Yang-Mills theory, as it enters the action by replacing $Q_\mu$ with $Q_\mu^\php + A^\zer_\mu$ in \eqref{gag2}. Therefore, we expect a gravitational interpretation of the finite-$\kappa$ Yang-Mills theory as in \cite{Steinacker:2010rh}. Since the matrix-type interactions are non-local, recovering Einstein gravity within this framework would only be possible through a local approximation of the dynamics for small $\kappa$. We are currently working towards this goal.

\subsection*{Acknowledgements}

Microsoft supported this research both by funding researchers and providing computational, logistical and other general resources. The authors thank Kevin Scott of Microsoft in particular for support of this project.

This research was supported in part by Perimeter Institute for Theoretical Physics. Research at Perimeter Institute is supported by the Government of Canada through Industry Canada and by the Province of Ontario through the Ministry of Research and Innovation. This research was also partly supported by grants from NSERC, FQXi and the John Templeton Foundation. V.S.~is supported by the Branco Weiss Fellowship - Society in Science, administered by
the ETH Zurich.

We thank Naghmeh Akhavan, Stephon Alexander, Rahul Balaji, Robert Brandenberger, Krish Desai, Michael Freedman, Edward Frenkel, Chetan Nayak, and Barbara Soda for valuable feedback and interactions.

\bibliography{References.bib}

\providecommand{\href}[2]{#2}\begingroup\raggedright\begin{thebibliography}{10}

\bibitem{Banks:1996vh}
T.~Banks, W.~Fischler, S.~H. Shenker, and L.~Susskind, ``{M theory as a matrix
  model: A Conjecture},''
  \href{http://dx.doi.org/10.1103/PhysRevD.55.5112}{{\em Phys. Rev. D}
  {\bfseries 55} (1997) 5112--5128},
  \href{http://arxiv.org/abs/hep-th/9610043}{{\ttfamily arXiv:hep-th/9610043}}.

\bibitem{Ishibashi:1996xs}
N.~Ishibashi, H.~Kawai, Y.~Kitazawa, and A.~Tsuchiya, ``{A Large N reduced
  model as superstring},''
  \href{http://dx.doi.org/10.1016/S0550-3213(97)00290-3}{{\em Nucl. Phys. B}
  {\bfseries 498} (1997) 467--491},
  \href{http://arxiv.org/abs/hep-th/9612115}{{\ttfamily arXiv:hep-th/9612115}}.

\bibitem{Taylor:2001vb}
W.~Taylor, ``{M(atrix) Theory: Matrix Quantum Mechanics as a Fundamental
  Theory},'' \href{http://dx.doi.org/10.1103/RevModPhys.73.419}{{\em Rev. Mod.
  Phys.} {\bfseries 73} (2001) 419--462},
  \href{http://arxiv.org/abs/hep-th/0101126}{{\ttfamily arXiv:hep-th/0101126}}.

\bibitem{Smolin:2000fr}
L.~Smolin, ``{The Cubic matrix model and a duality between strings and
  loops},'' \href{http://arxiv.org/abs/hep-th/0006137}{{\ttfamily
  arXiv:hep-th/0006137}}.

\bibitem{Smolin:2008pk}
L.~Smolin, ``{Matrix universality of gauge field and gravitational dynamics},''
  \href{http://arxiv.org/abs/0803.2926}{{\ttfamily arXiv:0803.2926 [hep-th]}}.

\bibitem{DiFrancesco:1993cyw}
P.~Di~Francesco, P.~H. Ginsparg, and J.~Zinn-Justin, ``{2-D Gravity and random
  matrices},'' \href{http://dx.doi.org/10.1016/0370-1573(94)00084-G}{{\em Phys.
  Rept.} {\bfseries 254} (1995) 1--133},
  \href{http://arxiv.org/abs/hep-th/9306153}{{\ttfamily arXiv:hep-th/9306153}}.

\bibitem{Ambjorn:2008jf}
J.~Ambjorn, R.~Loll, Y.~Watabiki, W.~Westra, and S.~Zohren, ``{A Matrix Model
  for 2D Quantum Gravity defined by Causal Dynamical Triangulations},''
  \href{http://dx.doi.org/10.1016/j.physletb.2008.06.026}{{\em Phys. Lett. B}
  {\bfseries 665} (2008) 252--256},
  \href{http://arxiv.org/abs/0804.0252}{{\ttfamily arXiv:0804.0252 [hep-th]}}.

\bibitem{Eichhorn:2013isa}
A.~Eichhorn and T.~Koslowski, ``{Continuum limit in matrix models for quantum
  gravity from the Functional Renormalization Group},''
  \href{http://dx.doi.org/10.1103/PhysRevD.88.084016}{{\em Phys. Rev. D}
  {\bfseries 88} (2013) 084016},
  \href{http://arxiv.org/abs/1309.1690}{{\ttfamily arXiv:1309.1690 [gr-qc]}}.

\bibitem{Alexander:2021rch}
S.~Alexander, W.~J. Cunningham, J.~Lanier, L.~Smolin, S.~Stanojevic, M.~W.
  Toomey, and D.~Wecker, ``{The Autodidactic Universe},''
  \href{http://arxiv.org/abs/2104.03902}{{\ttfamily arXiv:2104.03902
  [hep-th]}}.

\bibitem{Gross82}
D.~J. Gross and Y.~Kitazawa, ``{A quenched momentum prescription for large-N
  theories},'' {\em Nuclear Physics B} {\bfseries 206} no.~3, (1982) 440--472.

\bibitem{Kazakov:2000ar}
V.~A. Kazakov, ``{Field theory as a matrix model},''
  \href{http://dx.doi.org/10.1016/S0550-3213(00)00327-8}{{\em Nucl. Phys. B}
  {\bfseries 587} (2000) 645--656},
  \href{http://arxiv.org/abs/hep-th/0003065}{{\ttfamily arXiv:hep-th/0003065}}.

\bibitem{Yargic:2022ycw}
Y.~Yargic, J.~Lanier, L.~Smolin, and D.~Wecker, ``{A Cubic Matrix Action for
  the Standard Model and Beyond},''
  \href{http://arxiv.org/abs/2201.04183}{{\ttfamily arXiv:2201.04183
  [hep-th]}}.

\bibitem{Wigner32}
E.~Wigner, ``On the quantum correction for thermodynamic equilibrium,'' {\em
  Phys.~Rev.} {\bfseries 40} (1932) 749--759.

\bibitem{Weyl27}
H.~Weyl, ``{Quantenmechanik und Gruppentheorie},'' {\em Zeitschrift für
  Physik} {\bfseries 40} (1927) 1--46.

\bibitem{Groenewold46}
H.~J. Groenewold, ``On the principles of elementary quantum mechanics,'' {\em
  Physica} {\bfseries 12} no.~7, (1946) 405--460.

\bibitem{Moyal49}
J.~E. Moyal, ``Quantum mechanics as a statistical theory,'' {\em Mathematical
  Proceedings of the Cambridge Philosophical Society} {\bfseries 45} no.~1,
  (1949) 99–124.

\bibitem{Kontsevich:1997vb}
M.~Kontsevich, ``{Deformation quantization of Poisson manifolds. 1.},''
  \href{http://dx.doi.org/10.1023/B:MATH.0000027508.00421.bf}{{\em Lett. Math.
  Phys.} {\bfseries 66} (2003) 157--216},
  \href{http://arxiv.org/abs/q-alg/9709040}{{\ttfamily arXiv:q-alg/9709040}}.

\bibitem{Doplicher:1994tu}
S.~Doplicher, K.~Fredenhagen, and J.~E. Roberts, ``{The Quantum structure of
  space-time at the Planck scale and quantum fields},''
  \href{http://dx.doi.org/10.1007/BF02104515}{{\em Commun. Math. Phys.}
  {\bfseries 172} (1995) 187--220},
  \href{http://arxiv.org/abs/hep-th/0303037}{{\ttfamily arXiv:hep-th/0303037}}.

\bibitem{Douglas:2001ba}
M.~R. Douglas and N.~A. Nekrasov, ``{Noncommutative field theory},''
  \href{http://dx.doi.org/10.1103/RevModPhys.73.977}{{\em Rev. Mod. Phys.}
  {\bfseries 73} (2001) 977--1029},
  \href{http://arxiv.org/abs/hep-th/0106048}{{\ttfamily arXiv:hep-th/0106048}}.

\bibitem{Hull:2009mi}
C.~Hull and B.~Zwiebach, ``{Double Field Theory},''
  \href{http://dx.doi.org/10.1088/1126-6708/2009/09/099}{{\em JHEP} {\bfseries
  09} (2009) 099}, \href{http://arxiv.org/abs/0904.4664}{{\ttfamily
  arXiv:0904.4664 [hep-th]}}.

\bibitem{Aldazabal:2013sca}
G.~Aldazabal, D.~Marques, and C.~Nunez, ``{Double Field Theory: A Pedagogical
  Review},'' \href{http://dx.doi.org/10.1088/0264-9381/30/16/163001}{{\em
  Class. Quant. Grav.} {\bfseries 30} (2013) 163001},
  \href{http://arxiv.org/abs/1305.1907}{{\ttfamily arXiv:1305.1907 [hep-th]}}.

\bibitem{HOOFT1974461}
G.~Hooft, ``A planar diagram theory for strong interactions,''
  \href{http://dx.doi.org/https://doi.org/10.1016/0550-3213(74)90154-0}{{\em
  Nuclear Physics B} {\bfseries 72} no.~3, (1974) 461--473}.
  \url{https://www.sciencedirect.com/science/article/pii/0550321374901540}.

\bibitem{PhysRevLett.48.1063}
T.~Eguchi and H.~Kawai, ``Reduction of dynamical degrees of freedom in the
  large-$n$ gauge theory,''
  \href{http://dx.doi.org/10.1103/PhysRevLett.48.1063}{{\em Phys. Rev. Lett.}
  {\bfseries 48} (Apr, 1982) 1063--1066}.
  \url{https://link.aps.org/doi/10.1103/PhysRevLett.48.1063}.

\bibitem{Marino}
M.~Marino, {\em {Chern-Simons theory, matrix models, and topological strings}}.
\newblock Oxford: University Press, 2005.

\bibitem{Anninos:2020ccj}
D.~Anninos and B.~M\"uhlmann, ``{Notes on matrix models (matrix musings)},''
  \href{http://dx.doi.org/10.1088/1742-5468/aba499}{{\em J. Stat. Mech.}
  {\bfseries 2008} (2020) 083109},
  \href{http://arxiv.org/abs/2004.01171}{{\ttfamily arXiv:2004.01171
  [hep-th]}}.

\bibitem{Neumann1931}
J.~von Neumann, ``{Die Eindeutigkeit der Schrödingerschen Operatoren},'' {\em
  Math.~Ann.} {\bfseries 104} (1931) 570--578.

\bibitem{Zachos2017}
C.~K. Zachos, D.~B. Fairlie, and T.~L. Curtright, {\em {Quantum Mechanics in
  Phase Space}}.
\newblock World Scientific, 2005.

\bibitem{Steinacker:2010rh}
H.~Steinacker, ``{Emergent Geometry and Gravity from Matrix Models: an
  Introduction},'' \href{http://dx.doi.org/10.1088/0264-9381/27/13/133001}{{\em
  Class. Quant. Grav.} {\bfseries 27} (2010) 133001},
  \href{http://arxiv.org/abs/1003.4134}{{\ttfamily arXiv:1003.4134 [hep-th]}}.

\bibitem{Hillery84}
M.~Hillery, R.~O'Connell, M.~Scully, and E.~Wigner, ``Distribution functions in
  physics: Fundamentals,'' {\em Physics Reports} {\bfseries 106} no.~3, (1984)
  121--167.

\bibitem{Radicevic:2021ykf}
D.~Radicevic, ``{The Ultraviolet Structure of Quantum Field Theories. Part 1:
  Quantum Mechanics},'' \href{http://arxiv.org/abs/2105.11470}{{\ttfamily
  arXiv:2105.11470 [hep-th]}}.

\end{thebibliography}\endgroup
\bibliographystyle{Utphys}

\end{document}